# The object detection method aids in image reconstruction evaluation and clinical interpretation of meniscal abnormalities

Natalia Konovalova, MS[1]
Aniket Tolpadi, PhD[1,2]
Felix Liu, MS[1]
Zehra Akkaya, MD[1]
Felix Gassert, MD[1]
Paula Giesler, MD[1]
Johanna Luitjens, MD[1]
Misung Han, PhD[1]
Emma Bahroos, MS[1]
Sharmila Majumdar, PhD[1]
Valentina Pedoia, PhD[1]

[1] University of California, San Francisco
[2] University of California, Berkeley

University of California, San Francisco
BH203, Byers Hall, 600 16th St, San Francisco, CA 94158

**Corresponding author**: Natalia Konovalova
Email: natalia.konovalova@ucsf.edu
Address: San Francisco VA Medical Center, 4150 Clement St, San Francisco, CA 94121

**Funding**: This work was funded by NIH R01AR078762 (Apr 1, 2021 - Mar 31, 2026)

**Manuscript type**: Original Research

**Word count**: 3408 (including subtitles)

**Purpose**

To investigate the relationship between deep learning (DL) image reconstruction quality and anomaly detection performance and evaluate the efficacy of an artificial intelligence (AI) assistant in enhancing radiologists' interpretation of meniscal anomalies on reconstructed images.

**Materials and Methods**

In this retrospective study, an in-house reconstruction and anomaly detection pipeline was developed to assess knee MR images from 896 patients (mean age, 45; mean weight, 75 kg; 472 females). The original and 14 sets of DL-reconstructed images underwent evaluation using standard reconstruction and object detection metrics. Additionally, box-based reconstruction metrics were developed and compared to the standard evaluation approach. Two clinical radiologists conducted readings on a subset of 50 patients using both original and AI-assisted reconstructed images, with their accuracy and other performance characteristics subsequently assessed.

**Results**

Among the image-based reconstruction metrics, the structural similarity index (SSIM) demonstrated a weaker correlation with anomaly detection metrics (mAP, r=0.64, p=0.01; F1 score, r=0.38, p=0.18). Whereas, the box-based SSIM had a stronger association with detection performance (mAP, r=0.81, p<0.01; F1 score, r=0.65, p=0.01). Minor SSIM fluctuations did not impact detection outcomes, but substantial changes led to decreased performance. Radiologists' AI-assisted evaluations of reconstructed images showed improved accuracy (86.0% without assistance vs. 88.3% with assistance, p<0.05) and interrater agreement (Cohen's kappa, 0.39 without assistance vs. 0.57 with assistance). Following additional review, 17 more lesions were incorporated into the dataset.

**Conclusion**

The proposed anomaly detection method demonstrated promise in both evaluating reconstruction algorithms for automated downstream tasks and aiding radiologists in interpreting DL-reconstructed MR images.

## Abbreviations

MRI – magnetic resonance imaging, AI – artificial intelligence, DL – deep-learning, CNN – convolutional neural network, FSE – fast spin echo, CUBE - cube ultra-fast balanced echo, SSIM – structural similarity index, PSNR – peak signal-to-noise ratio, nRMSE – normalized root mean squared error, mAP – mean average precision.

# Introduction

Meniscal lesions represent a prevailing concern among individuals with knee problems and constitute the leading cause of orthopedic surgical interventions in the United States (1). Injuries to meniscal tissues have been linked to the progression of osteoarthritis and vice versa (2). While treatment outcomes are often satisfactory, many patients still report reduced knee mobility compared to the general population (3). These factors impose a substantial financial burden on medical systems (4) and underscore the need for accurate diagnosis of meniscal anomalies to ensure effective clinical management and favorable patient prognosis.

MRI remains one of the most effective non-invasive tests for detecting meniscal anomalies, offering high-resolution images and excellent soft tissue contrast (5). In recent years, deep learning (DL) techniques have revolutionized various aspects of medical imaging, including acquisition, post-processing, and analysis. As such, CNNs, along with transformers and diffusion models, exhibited remarkable capabilities in accelerating MR image reconstruction from significantly under-sampled data (6–9). Simultaneously, segmentation and object detection algorithms demonstrated considerable potential in advancing clinical decision-making (10,11).

Despite extensive efforts to integrate DL-reconstructed images into clinical practice, limited research has focused on assessing their usability for alternative downstream tasks (12,13). Traditional evaluation metrics for image reconstruction, such as normalized root mean square error (nRMSE), peak signal-to-noise ratio (PSNR), and structural similarity index (SSIM), primarily aim to produce high-quality images for visual inspection by radiologists (14–16). However, there is currently no clear evidence to suggest that optimizing these metrics results in the best datasets for training detection or segmentation models. Moreover, this evaluation approach has revealed specific limitations, particularly concerning knee imaging, where top-performing reconstruction models, based on standard reconstruction metrics, have failed to preserve essential fine features and small lesions within the meniscal area (17).

In response to these challenges, our feasibility study investigates the relationship between image reconstruction and object detection performance by employing widely accepted metrics to measure their correlation and impact. We hypothesize that anomaly detection can serve as a vital tool for evaluating the performance of DL-based reconstruction models, particularly when the intention is to use reconstructed images for subsequent automated tasks. We also emphasize the benefits of our anomaly detection AI assistant in enhancing radiologists' performance when interpreting reconstructed images for clinical practice. Additionally, we aim to evaluate the effectiveness of object detection in assessing the extent to which image reconstruction techniques preserve crucial information within MR images. To achieve this, we calculate nRMSE, PSNR, and SSIM exclusively within predicted regions of interest, defined by bounding boxes, and compare them to standard reconstruction metrics.

## Materials and Methods

**Dataset**

This retrospective study was conducted in accordance with a process approved by the local institution review board with waived informed consent. In this study, we collected a dataset of knee MRI exams from the local clinical population between June 2021 and June 2022. These patients presented a variety of knee abnormalities, including bone, cartilage, and meniscal lesions, anterior and posterior cruciate ligament (ACL and PCL) tears, and ACL-reconstructed knees. No exclusion criteria were applied upon selection.

**MRI Acquisition**

3D fast spin-echo (FSE) fat-suppressed CUBE images were acquired at a GE Discovery MR750 scanner using 18-channel knee transmit/receive coil with the following parameters: repetition time (TR)/echo time (TE), 1002/29 msec; field of view (FOV), 15 cm$^2$; acquisition matrix, 256×256×200; slice thickness, 0.6 mm; echo train length, 36; readout bandwidth, ±62.5 kHz; acceleration, 4X ARC (18); acquisition time, 4 min 58 sec. Subsequently, an in-house pipeline was developed that leveraged GE Orchestra 1.10 and other post-processing tools to reconstruct images from ARC-undersampled k-space data and store them as DICOM files with uniform matrix dimensions of 512x512. ARC-reconstructed multicoil k-space data was also saved.

**Annotation**

The data was anonymized by removing patient-sensitive information from the DICOM headers. Annotation was performed using an online platform (MD.ai, New York, NY). Three radiologists (J.L. with 2 years of training, F.G. and P.G. both with 3 years of training) manually marked all knee anomalies by drawing bounding boxes on each sagittal image slice, as shown in Figure 1. To calibrate the annotation procedure, the radiologists initially evaluated 15 cases together. The readers were then assigned nonoverlapping exams and instructed to specify the anatomical location of the pathology when labeling the bounding boxes. This study focused on lesions in six meniscal compartments: the medial and lateral meniscal horns and bodies. The labels served as the ground truth for anomaly detection evaluation. The radiologists also identified cases with insufficient image quality for annotation. A full list of labels and counts is in Table E1 (supplement).

**Anomaly Detection Pipeline**

An automated anomaly detection pipeline was implemented with Python version 3.8 (Python Software Foundation), PyTorch version 1.12.1 (https://pytorch.org/), and PyTorch Lightning version 1.7.7 (https://lightning.ai/). A Faster R-CNN model with a ResNet-50-FPN pretrained backbone (19) from PyTorch Torchvision version 0.13.1 was used for detection on 2D image slices. As a preprocessing step, pixel data was extracted from DICOM volumes, image intensities were normalized by percentile-based normalization, and all images were saved as sagittal-oriented 2D arrays. Bounding boxes were converted to the PASCAL VOC format (20). These constituted the inputs to the object detection model. All meniscal anomaly labels were considered a single class for training purposes.

The dataset was split into 80% training, 10% validation, and 10% testing partitions on the patient level, ensuring no data leaked between the splits. To enhance training efficiency, a data fractionation approach was implemented, where each epoch was trained on 20% of randomly selected slices from the whole dataset. The data augmentation protocol was based on previously published research and included a custom implementation of a bounding box bidirectional shift function among other standard techniques (21). Additionally, since the inherent variability in the number of different meniscal anomalies existed in our dataset, a bounding box upsampling technique was utilized to counteract the class imbalance within the anomaly class. The model used a Stochastic Gradient Descent (SGD) optimizer with a learning rate scheduler and an initial learning rate of 0.01. The training was conducted on two Tesla V100 32GB NVIDIA GPUs for a maximum of 30 epochs. Detection performance was assessed by precision, recall, mean average precision (mAP), and F1 score. True positive predictions were defined as those with at least 0.2 Intersection-over-Union (IoU) and a 0.7 confidence score. A detailed description of data fractionation, data augmentation, bounding box upsampling, and the full list of adjustable training parameters are provided in Appendix E1 and Table E2 (supplement).

**Detection Evaluation on Reconstructed Images**

To assess the performance of anomaly detection on reconstructed images, we created 14 additional test sets using an in-house pipeline featuring 8X accelerated DL image reconstruction (22). Each set included reconstructed images for the same patients as in the original detection test set, representing the inference results of KIKI-inspired I-Net (23) and similar architecture UNet models trained with specific combinations of common loss functions. All UNet and I-Net configurations were trained for 15 epochs with a learning rate of 0.001. We also generated zero-filled and fully sampled sets for comparison. All test sets were normalized as required by the anomaly detection pipeline before inference, ensuring consistent image quality. More details on sampling patterns and reconstruction are provided in Figure 2.

      We calculated nRMSE, PSNR, and SSIM in two ways: the standard method based on the entire 3D image volume and an alternative method based only on pixels within predicted bounding boxes, with the calculation algorithm detailed in Figure E1 (supplement). Object detection metrics were also calculated for all reconstructed sets to explore potential correlations with reconstruction metrics. To evaluate the possible link between reconstruction performance and object detection outcomes, we compared the mean values of SSIM, PSNR, and nRMSE based on slices where true positive (TP), false positive (FP), and false negative (FN) predictions were observed. Additionally, we calculated Spearman's

correlation coefficients with corresponding p-values between the boxes' confidence scores and each of the box-based reconstruction metrics.

**AI-assisted Reading**

To assess the impact of an AI assistant on reading results, two radiologists (Z.A. with 10 years of clinical experience and J.L.) each examined 30 cases from the testing partition. They were tasked with identifying pathologic or non-pathologic conditions in the six meniscal compartments under four conditions: original DICOM images (ground truth), reconstructed images without AI assistance, and original and reconstructed images with AI assistance using predicted anomaly boxes. The selection of the reconstructed image set for this analysis was based on the highest SSIM score. Readings were conducted using MD.ai, with a two-week wash-out period between experiments. Additionally, an expert radiologist (Z.A.) compared the results of readings performed on original images with and without boxes to identify any anomalies that may have been missed during the initial ground truth annotation. This experiment was conducted more than a year after the initial dataset annotation by J.L., which minimized the reader's bias.

**Relationship Between Image Quality and Detection Performance**

To comprehensively evaluate the relationship between image reconstruction and object detection performance, we devised eight distinct test sets by introducing controlled alterations to image quality. Utilizing a random number generator, we generated noise with specified magnitudes, denoted by various noise constants (e.g., 0.05, 0.10, 0.15, 0.20). By randomly changing the noise sign for specific pixels, we produced diverse noise patterns. We then integrated this noisy data into the ground truth image, resulting in two variations for each noise level: one with both positive and negative perturbations (referred to as "Noise x2.0," and similar designations), and the other with solely positive perturbations (known as "Const x2.0," and so on). Notably, this approach allowed us to maintain nRMSE and PSNR as constants for each noise increment while achieving substantial variations in SSIM only (24). Our primary goal was to investigate how large changes in SSIM, as a metric of image quality, correlate with object detection performance.

**Statistical Analysis**

We conducted statistical analysis using SciPy version 1.9.3 (https://scipy.org/) and scikit-learn version 1.2.2 (https://scikit-learn.org/). Spearman's correlation, along with its corresponding p-value, was employed to assess metric relationships. Additionally, one-way ANOVA was performed to compare the means of metrics' distributions, followed by a post hoc pairwise t-test to determine any deviating groups. We utilized the $Chi^2$ contingency test to determine the significance of AI assistance in radiologists' readings and Cohen's Kappa to evaluate interrater reliability (25).

## Results

**Dataset Characteristics**

A total of 947 knee MRI exams were obtained and 51 were excluded due to poor image quality, leaving a dataset of 896 exams (175,492 slices). Of those, 406 patients had at least one meniscal abnormality, yielding a total of 18,059 bounding boxes drawn, and others had healthy menisci. The mean age was 44.7 ± 15.3 years, the mean weight was 74.8 ± 15.8 kg, and 52.7% (472 of 896) were females. Additional demographic characteristics of data partitions are summarized in Table 1.

**Anomaly Detection and Reconstruction Performance**

The anomaly detection model achieved the following results on original images: 70.53% precision, 72.17% recall, 63.09% mAP, and a 71.34% F1 score. For a summary and examples of its performance on reconstruction test sets, refer to Table 2 and Figure 3. When employing the classic L1-loss function, the UNet model demonstrated superior nRMSE, PSNR, and SSIM values. Combining L1 and SSIM losses yielded similar performance. Faster R-CNN performed well on reconstructed test sets, revealing strong correlations between box-based nRMSE, PSNR, and standard reconstruction metrics (r = 1.00, p < 0.05 for both metrics). Box-based SSIM exhibited a robust correlation with its standard counterpart (r = 0.76, p < 0.05). A heatmap displayed in Figure 4 illustrates the relationship between anomaly detection and reconstruction performance, highlighting moderate to strong correlations between mAP, F1 score, and nRMSE/PSNR. Importantly, image-based SSIM showed only a moderate or insignificant correlation with detection metrics (mAP: r = 0.64, p < 0.05; F1: r = 0.38, p > 0.05), while box-based SSIM exhibited a strong correlation with detection metrics (mAP: r = 0.81, p < 0.05; F1: r = 0.65, p < 0.05) (26). Additionally, the comparison of mean values of SSIM, PSNR, and nRMSE calculated separately for slices with TP, FP, and FN predictions did not reveal any significant patterns that would suggest a strong predictor of detection performance. An example of this analysis is presented in Figure 5, with plots for other reconstruction models depicted in Figure E2 (supplement). No significant correlation was observed between image-based reconstruction metrics and prediction confidence scores, as demonstrated in Table 3.

**AI-assisted Reading Results**

The reconstructed images from the UNet with L1 loss were selected for assessment because the model showed top results in both reconstruction and detection evaluation. In total, 50 studies were reviewed, with 10 overlapping to assess interrater variability. Random stratification ensured diverse predictions in each set of 30 studies: 16 true positives with at least one lesion per patient, 6 false positives, and 8 anomaly-free cases. Reading without AI assistance resulted in fair interrater agreement, as measured by Cohen's Kappa, for both the original DICOM image set (k = 0.41) and the reconstructed image set (k = 0.39). However, the addition of anomaly boxes predicted by an AI assistant significantly increased agreement for both sets of readings, resulting in k = 0.60 for the original image set and k = 0.57 for the reconstructed image set, respectively. Furthermore, the overall performance of the radiologists, as measured by accuracy, precision, recall, specificity, and F1 score improved (p-values < 0.05) when reading the reconstructed images assisted with boxes, as illustrated in Table 4. The comparison of results from reading original DICOM images with and without AI-predicted boxes led to the reclassification of false positive cases, which were not reported during the initial annotation, as true positives and the subsequent addition of 17 new lesions to the dataset.

**Observations on the Influence of Image Quality on Detection Performance**

Adding noise perturbations to images resulted in a gradual decrease in image quality. Within the same noise increment, adding both random positive and negative noise resulted in worse image quality than adding just positive values (termed "Const"), as measured by SSIM, while nRMSE and PSNR remained unchanged for image-based metrics, as shown in Table E3 (supplement). For box-based metrics, achieving the same nRMSE and PSNR values within one noise increment was impossible, but the values were still very close. Object detection performance, measured by recall, mAP, and F1 score, decreased proportionally to the decrease in image quality measured by standard reconstruction metrics (Spearman's correlation coefficients were 1.00 and p-values were 0.00 between each of the detection and reconstruction metrics), as shown in Figure 6. However, precision did not follow this trend (Spearman's correlation coefficients were 0.40 and p-values were 0.60 between precision and each of the reconstruction metrics). It is worth noting that the impact on detection performance seemed to decrease as image quality improved, despite similar variations in SSIM within one noise increment.

## Discussion

In this study, we sought to address ongoing challenges in assessing the quality of DL-reconstructed images and their utility for alternative downstream tasks, as well as their visual assessment by trained radiologists. We began with the premise that models optimized for the visual interpretation of clinical images might not be ideally suited as inputs for other image processing tasks, like segmentation or anomaly detection. Nevertheless, with the increasing integration of AI into clinical practice, there is a growing need for accelerated reconstruction methods that produce images fit for further automated analysis (27–29). Our research examines the relationship between image reconstruction and anomaly detection, specifically within the context of meniscal anomalies, aiming to enhance the assessment of reconstruction quality and to evaluate the utility of the detector for routine clinical interpretation of reconstructed images.

The correlation analysis between structural similarity (SSIM) and object detection metrics emphasizes the value of anomaly detection as an additional tool for evaluating image reconstruction models, especially when further image analysis is intended. Image-based SSIM was found to have a weaker correlation with detection metrics compared to box-based SSIM. This indicates that image-based

SSIM may not be the most reliable indicator for preserving vital information in reconstructed images for anomaly detection purposes, such as small lesion details, while box-based SSIM may offer more utility. This is in line with other research that also reported a lack of correlation between image-based SSIM and detection metrics (30). Significantly, it was observed that the best-performing reconstruction models, such as UNet trained with L1 or a combination of L1 and SSIM loss functions, did not always lead to the best object detection performance, further implying that conventional reconstruction metrics may not be sufficient for choosing reconstruction models for advanced automated tasks.

To further substantiate the importance of integrating anomaly detection into the image reconstruction evaluation process, we analyzed mean image-based SSIM values separated into true positive (TP), false positive (FP), and false negative (FN) prediction groups. Our results did not reveal any significant patterns that suggest an increase in TP predictions associated with higher SSIM. In some models, we observed significant differences in FN with lower or higher mean SSIM. Overall, our experiments showed that small SSIM variations did not appear to strongly impact detection predictions. However, the absence of a clear-cut link between image quality, as measured by SSIM and other conventional reconstruction metrics, and detection performance, should not be interpreted as suggesting that these metrics are unrelated.

Our investigation demonstrated that changes in image quality, particularly in SSIM, did not proportionally affect detection performance. When we introduced noise, while keeping nRMSE and PSNR constant, we observed that detection performance was significantly impacted by larger negative shifts in SSIM, leading to poorer detection results. Conversely, as image quality improved, its effect on detection accuracy diminished. This suggests that while substantial SSIM alterations can influence downstream task performance, minor SSIM fluctuations are less consequential. Additionally, this indicates that high SSIM scores might correlate with a plateau in detection sensitivity to SSIM changes. These insights highlight the necessity of developing specialized metrics for assessing image quality tailored to specific downstream tasks, implying that achieving peak performance as per conventional reconstruction metrics may not be necessary for reconstruction model optimization.

Another discovery from our research is the beneficial impact of AI-assisted anomaly detection on the performance of radiologists interpreting reconstructed images. The use of AI-predicted anomaly boxes significantly increased interrater agreement among radiologists and improved their performance in terms of accuracy, precision, recall, specificity, and F1 score. Implementing AI tools like anomaly detection algorithms is crucial in effectively integrating DL-based image reconstruction into clinical workflows. This may enhance the diagnostic accuracy and consensus among radiologists, which can be variable even among seasoned practitioners (31). Additionally, these tools can facilitate dataset annotation and instructional processes for radiology trainees (32–34).

Our study has several limitations. We focused narrowly on knee MRI examinations featuring meniscal anomalies, which are anatomically distinct. Thus, our findings might not extend as effectively to different anatomical areas or other types of anomalies, like cartilage abnormalities. There is a need for future research to test the wider applicability of our results in various clinical contexts. We also specifically employed a 3D fast spin-echo fat-suppressed CUBE MRI sequence that, despite its effectiveness in our research, might not be the most common in diverse clinical settings (35). A broader validation of our findings across various knee MRI protocols would be beneficial. In addition, our study relied on older reconstruction models to integrate image reconstruction with detection pipelines. Newer methodologies, such as variational networks and transformers, warrant investigation for potentially deeper insights (6,36). Recent literature also indicates that the loss of fine but clinically significant features could stem from the chosen undersampling technique, not solely the reconstruction algorithm, as different random sampling seeds can influence the performance of reconstruction models (37). This underlines the need for further analysis to determine the practicality of the anomaly detection approach under these conditions.

Our feasibility study underscores the critical role of anomaly detection in the assessment of DL-based image reconstruction models, particularly when aiming to maintain clinically significant features. We determined that image-based SSIM is not necessarily indicative of detection performance, highlighting a complex relationship between image quality as measured by SSIM and detection outcomes.

Incorporating anomaly detection and box-based reconstruction metrics is crucial for evaluating reconstruction models for downstream applications. Moreover, our research points to the beneficial effects of AI-assisted detection on the interpretive accuracy of radiologists, thereby expanding the potential for clinical use. Looking ahead, we suggest investigating reconstruction and anomaly detection pipelines that operate concurrently, using the detector as a penalizing factor, and exploring the use of object detection to refine undersampling patterns to optimize reconstruction results.

Table 1: Patient Demographics and Data Partitioning

| Parameter | Training | Validation | Testing |
|---|---|---|---|
| Demographic characteristics | | | |
|     Number of patients | 716 | 90 | 90 |
|     Number of females | 376 | 48 | 48 |
|     Mean age (y) | 44.2 ± 15.5 | 48.1 ± 14.8 | 44.7 ± 14.1 |
|     Mean weight (kg) | 75.3 ± 15.9 | 72.7 ± 14.6 | 73.4 ± 16.0 |
| Anomaly distribution | | | |
|     Number of slices | 140,202 | 17,640 | 17,650 |
|     Anomaly boxes | 14,550 | 1,714 | 1,795 |
|     Slices with anomaly boxes | 12,857 | 1,533 | 1,590 |

Note. Age and weight are calculated as mean ± standard deviation. Data partitions were 80% training, 10% validation, and 10% testing.

Table 2: Detection and Reconstruction Performance

| # | Model Name | Recon Image-Based Metrics | | | Detection Metrics | | | | Recon Boxes-Based Metrics | | |
|---|---|---|---|---|---|---|---|---|---|---|---|
| | | nRMSE | PSNR | SSIM | Precision | Recall | mAP | F1 | nRMSE | PSNR | SSIM |
| 1 | Zero-Filled | 0.45 ± 0.12 | 24.12 ± 1.97 | 29.36 ± 5.38 | 83.80 | 42.04 | 39.65 | 55.99 | 0.45 ± 0.12 | 12.71 ± 2.34 | 29.99 ± 10.67 |
| 2 | UNet (K-Space) | 0.08 ± 0.05 | 31.84 ± 2.76 | 72.61 ± 4.69 | 61.83 | 73.06 | 61.12 | 66.98 | 0.08 ± 0.04 | 20.86 ± 2.47 | 67.13 ± 9.62 |
| 3 | UNet (L1) | **0.02 ± 0.01** | **36.75 ± 1.97** | **81.63 ± 2.59** | **69.79** | **71.06** | **61.48** | **70.42** | **0.03 ± 0.02** | **24.48 ± 2.18** | **79.65 ± 5.79** |
| 4 | UNet (SSIM) | 0.05 ± 0.02 | 34.16 ± 1.90 | 78.82 ± 2.78 | 67.67 | 66.29 | 56.92 | 66.97 | 0.05 ± 0.02 | 22.10 ± 2.19 | 76.50 ± 5.44 |
| 5 | UNet (L1, K-Space) | 0.20 ± 0.09 | 27.93 ± 3.97 | 62.91 ± 8.54 | 71.96 | 66.43 | 58.84 | 69.08 | 0.20 ± 0.09 | 16.61 ± 2.93 | 57.53 ± 11.42 |
| 6 | UNet (K-Space, SSIM) | 0.07 ± 0.04 | 33.09 ± 3.23 | 75.02 ± 4.79 | 69.45 | 70.16 | 60.60 | 69.80 | 0.07 ± 0.04 | 21.30 ± 2.88 | 75.54 ± 7.18 |
| 7 | UNet (L1, SSIM) | **0.04 ± 0.02** | **35.35 ± 2.28** | **81.27 ± 2.62** | **67.81** | **71.47** | **61.43** | **69.59** | **0.04 ± 0.02** | **22.84 ± 2.25** | **79.64 ± 5.22** |
| 8 | UNet (L1, K-Space, SSIM) | 0.55 ± 0.10 | 23.09 ± 2.17 | 38.25 ± 7.48 | 83.63 | 49.94 | 46.02 | 62.54 | 0.52 ± 0.08 | 11.85 ± 2.00 | 25.34 ± 6.77 |
| 9 | I-Net (K-Space) | 0.26 ± 0.13 | 27.04 ± 4.21 | 58.02 ± 10.48 | 68.92 | 65.26 | 56.12 | 67.04 | 0.26 ± 0.10 | 15.66 ± 2.90 | 47.37 ± 12.02 |
| 10 | I-Net (L1) | 0.03 ± 0.01 | 36.08 ± 2.07 | 74.43 ± 3.37 | 69.81 | 70.03 | 60.66 | 69.92 | 0.03 ± 0.01 | 24.21 ± 2.12 | 78.81 ± 5.94 |
| 11 | I-Net (SSIM) | 0.04 ± 0.02 | 34.54 ± 2.57 | 77.15 ± 3.10 | **70.20** | **70.61** | **61.54** | **70.41** | 0.05 ± 0.02 | 22.40 ± 2.12 | 77.89 ± 4.88 |
| 12 | I-Net (L1, K-Space) | 0.03 ± 0.01 | 35.81 ± 2.44 | 76.54 ± 3.47 | 70.85 | 68.30 | 59.52 | 69.55 | 0.03 ± 0.01 | 24.03 ± 1.98 | 77.85 ± 5.94 |
| 13 | I-Net (K-Space, SSIM) | 0.17 ± 0.11 | 29.32 ± 4.89 | 65.05 ± 9.27 | 73.88 | 66.68 | 58.51 | 70.10 | 0.16 ± 0.08 | 17.65 ± 3.63 | 64.94 ± 10.28 |
| 14 | I-Net (L1, SSIM) | 0.03 ± 0.01 | 35.44 ± 2.34 | 72.25 ± 3.81 | 68.42 | 71.08 | 61.02 | 69.72 | 0.04 ± 0.01 | 23.40 ± 2.05 | 79.15 ± 5.11 |
| 15 | I-Net (L1, K-Space, SSIM) | 0.03 ± 0.01 | 35.66 ± 2.52 | 73.82 ± 3.63 | **69.80** | **71.21** | **61.18** | **70.50** | 0.04 ± 0.01 | 23.58 ± 2.08 | 79.44 ± 5.00 |
| 16 | Fully-Sampled | - | - | - | 74.70 | 66.88 | 59.29 | 70.57 | - | - | - |

Note. nRMSE – normalized root mean squared error. PSNR – peak signal-to-noise ratio. SSIM – structural similarity index, reported in %. All detection metrics are reported in %. mAP – mean average precision. F1 – F1 score. Fully-sampled detection results are provided for reference. Results of best-performing models in terms of reconstruction and detection outcomes are in bold.

Table 3: Spearman's Correlation between Reconstruction Metrics and Prediction Confidence Scores

| # | Model Name | True Positive (TP) Predictions | | | False Positive (FP) Predictions | | |
|---|---|---|---|---|---|---|---|
| | | SSIM | PSNR | nRMSE | SSIM | PSNR | nRMSE |
| 1 | Zero-Filled | 0.22 (0.00) | 0.04 (0.25) | -0.14 (0.00) | -0.03 (0.63) | 0.16 (0.01) | -0.01 (0.92) |
| 2 | UNet (K-Space) | 0.21 (0.00) | 0.01 (0.68) | -0.06 (0.02) | 0.17 (0.00) | 0.06 (0.04) | -0.11 (0.00) |
| 3 | UNet (L1) | 0.14 (0.00) | -0.01 (0.59) | -0.10 (0.00) | 0.03 (0.48) | 0.01 (0.80) | -0.06 (0.07) |
| 4 | UNet (SSIM) | 0.08 (0.01) | -0.06 (0.03) | -0.01 (0.72) | -0.10 (0.01) | -0.04 (0.21) | 0.04 (0.29) |
| 5 | UNet (L1, K-Space) | 0.03 (0.25) | -0.03 (0.38) | 0.04 (0.16) | 0.19 (0.00) | 0.19 (0.00) | -0.21 (0.00) |
| 6 | UNet (K-Space, SSIM) | 0.21 (0.00) | 0.05 (0.07) | -0.09 (0.00) | 0.03 (0.48) | -0.04 (0.20) | 0.01 (0.80) |
| 7 | UNet (L1, SSIM) | 0.16 (0.00) | -0.08 (0.01) | -0.01 (0.83) | 0.04 (0.23) | 0.04 (0.25) | -0.07 (0.06) |
| 8 | UNet (L1, K-Space, SSIM) | 0.30 (0.00) | 0.06 (0.09) | -0.27 (0.00) | 0.24 (0.00) | 0.22 (0.00) | -0.20 (0.00) |
| 9 | I-Net (K-Space) | 0.14 (0.00) | 0.02 (0.57) | -0.03 (0.32) | 0.14 (0.00) | 0.07 (0.05) | -0.13 (0.00) |
| 10 | I-Net (L1) | 0.23 (0.00) | 0.10 (0.00) | -0.20 (0.00) | 0.02 (0.56) | -0.03 (0.49) | -0.02 (0.50) |
| 11 | I-Net (SSIM) | 0.12 (0.00) | -0.07 (0.01) | 0.08 (0.01) | -0.03 (0.41) | -0.08 (0.03) | 0.06 (0.10) |
| 12 | I-Net (L1, K-Space) | 0.19 (0.00) | 0.03 (0.31) | -0.12 (0.00) | 0.04 (0.34) | -0.03 (0.39) | -0.03 (0.47) |
| 13 | I-Net (K-Space, SSIM) | -0.01 (0.82) | -0.07 (0.02) | 0.09 (0.00) | 0.13 (0.00) | 0.14 (0.00) | -0.13 (0.00) |
| 14 | I-Net (L1, SSIM) | 0.21 (0.00) | 0.03 (0.28) | -0.11 (0.00) | -0.04 (0.22) | -0.10 (0.01) | 0.11 (0.00) |
| 15 | I-Net (L1, K-Space, SSIM) | 0.15 (0.00) | -0.06 (0.02) | 0.07 (0.01) | -0.01 (0.76) | -0.09 (0.01) | 0.13 (0.00) |

Note. SSIM – structural similarity index, reported in %. PSNR – peak signal-to-noise ratio. nRMSE – normalized root mean squared error. All numbers are in the form: "correlation coefficient (p-value)."

Table 4: AI-Assisted Reading Results

| Metric | Without Boxes | With Boxes |
| --- | --- | --- |
| Accuracy | 86.00 | 88.33 ↑ |
| Precision | 65.39 | 70.67 ↑ |
| Recall | 77.27 | 80.30 ↑ |
| Specificity | 88.46 | 90.60 ↑ |
| F1 | 70.83 | 75.18 ↑ |
| Cohen's Kappa | 0.39 | 0.57 ↑ |
| p-value | < 0.05 | < 0.05 |

Note. All performance values are reported in %. The statistical significance of AI-assisted reading results was calculated using the $Chi^2$ contingency test.

Figure 1: Annotations

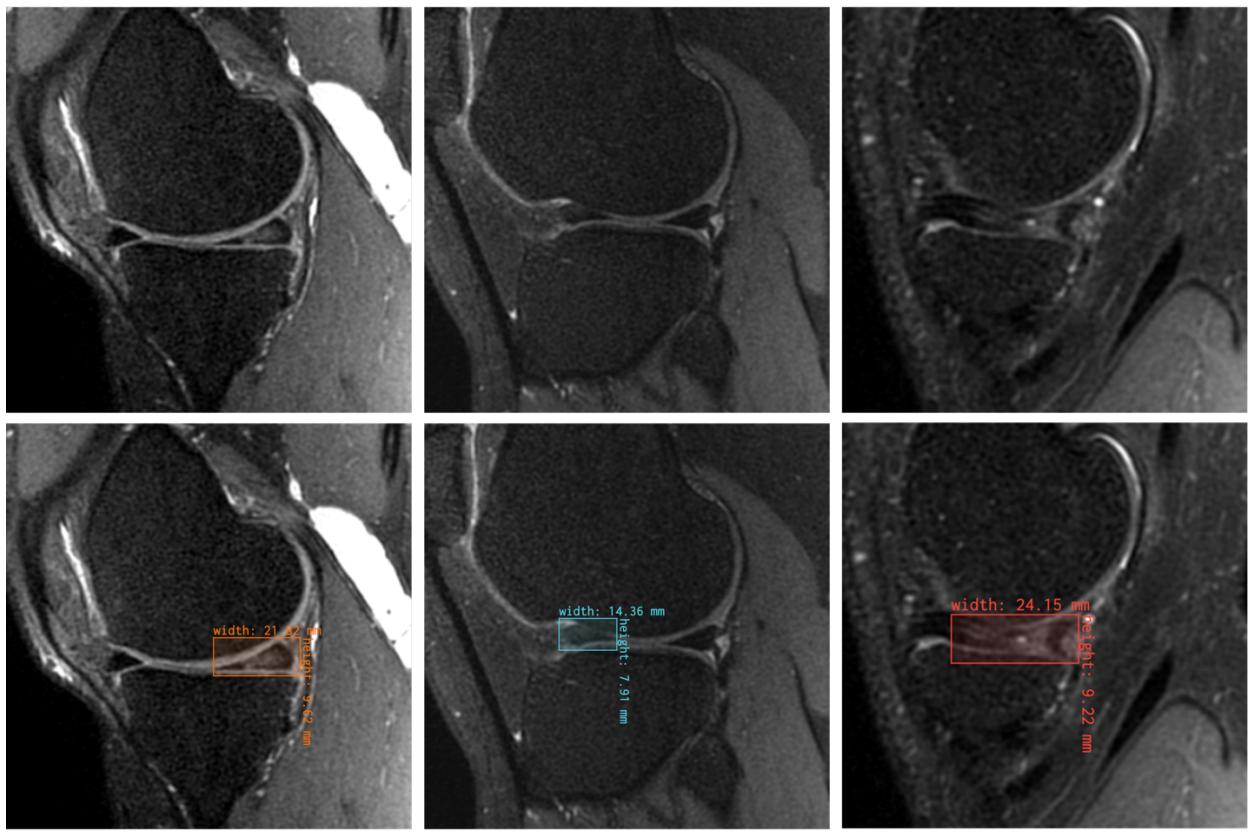

Examples of manual annotations created by radiologists using the MD.ai platform. Top row: no annotations, bottom row: corresponding boxes. From left to right: lesions in the medial posterior horn, lateral anterior horn, and medial meniscal body.

Figure 2: Anomaly Detection and Reconstruction Pipeline

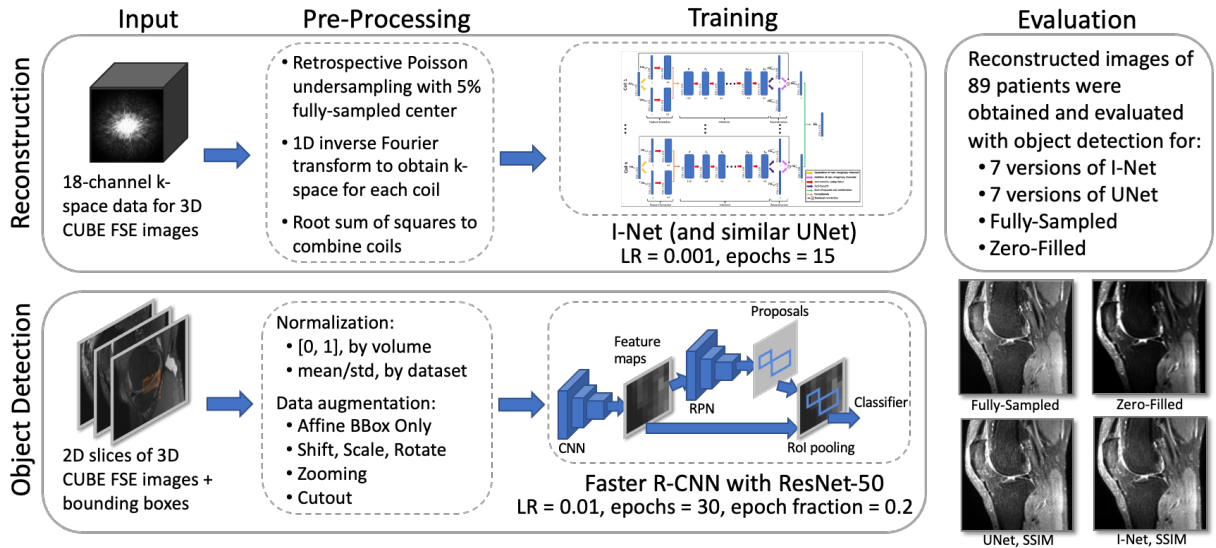

Reconstruction and detection pipelines were curated in parallel, and the shared test set of 89 patients was used for evaluation and analysis. For reconstruction, ARC-reconstructed 3D multicoil k-space was 8X undersampled in $k_y$-$k_z$ with a center-weighted Poisson pattern while fully sampling the 5% central square in k-space. Undersampled and fully-sampled k-space were 1D inverse Fourier transformed along $k_z$ direction, yielding undersampled and corresponding ground truth $k_x$-$k_y$-z 2D k-space for each coil. Root sum of squares coil combination of fully sampled coil images was used to calculate ground truth coil-combined images. KIKI-style I-Nets and UNets were trained with the following loss functions: (1) multi-coil k-space loss; (2) coil-combined L1 loss; (3) coil-combined SSIM loss; (4) coil-combined L1 and multi-coil k-space loss; (5) multi-coil k-space and coil-combined SSIM loss; (6) coil-combined L1 and SSIM loss; (7) coil-combined L1 and SSIM and multi-coil k-space loss. For object detection, 2D slices were obtained from 3D FSE CUBE images, and normalization and data augmentation were applied before entering the training cycle.

Figure 3: Example of Reconstruction and Anomaly Detection for UNet and I-Net

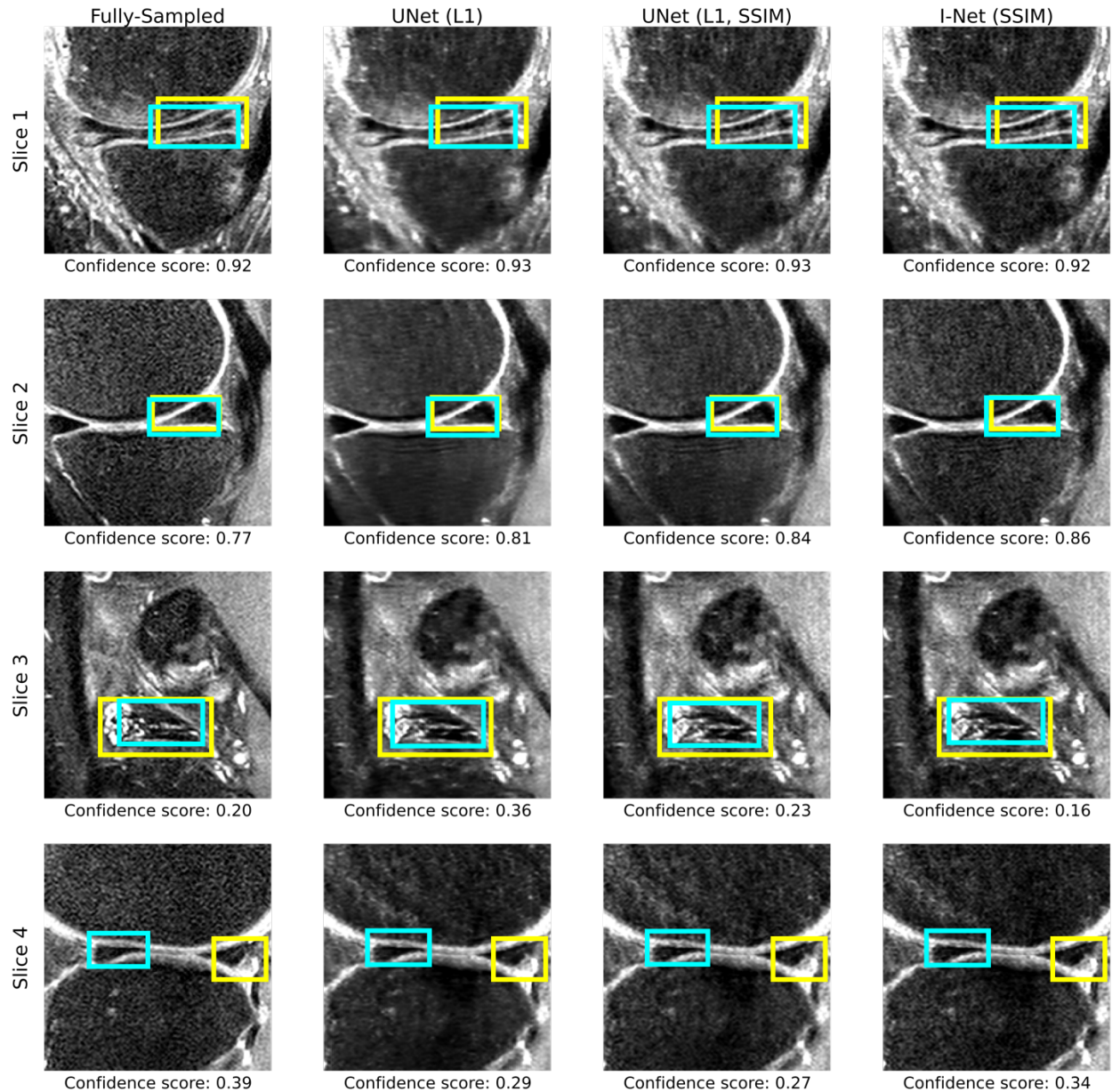

Rows represent four different sagittal slices; columns show fully-sampled and top three reconstruction models' outcomes. Ground truth boxes are in yellow, and predicted boxes are in blue. Slice 1: the lesion is visible, and the detection model yielded high-confidence true positive predictions. Slice 2: the lesion is less visible, and predictions are still within the confidence threshold of 0.75 and considered true positive. Slice 3: the lesion is visible; however, the model failed to predict with high confidence. Slice 4: the detection model fails to predict the lesion's correct location.

Figure 4: Detection and Reconstruction Metrics Correlation

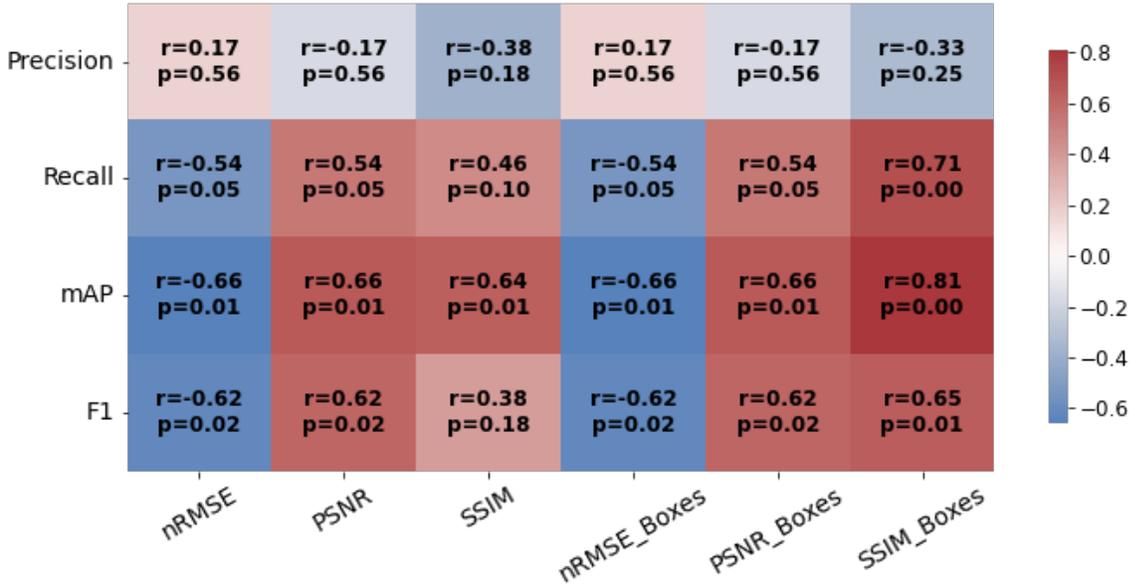

Spearman's correlations with corresponding p-values between standard and boxes-based reconstructions and object detection metrics are presented. Both classic and boxes-based nRMSE and PSNR exhibit moderate negative and positive correlations, respectively, with the mAP and F1 scores ($p < 0.05$). Importantly, classic SSIM shows a moderate correlation with mAP and a low, insignificant ($p > 0.05$) correlation with the F1 score, while boxes-based SSIM demonstrates a strong positive correlation with mAP and a moderate correlation with the F1 score. These findings support the hypothesis that the boxes-based metric is a more indicative measure of specific downstream task performance. It may, therefore, serve as an additional tool for evaluating reconstruction performance.

Figure 5: Relationship Between Standard Reconstruction Metrics and Detection Performance

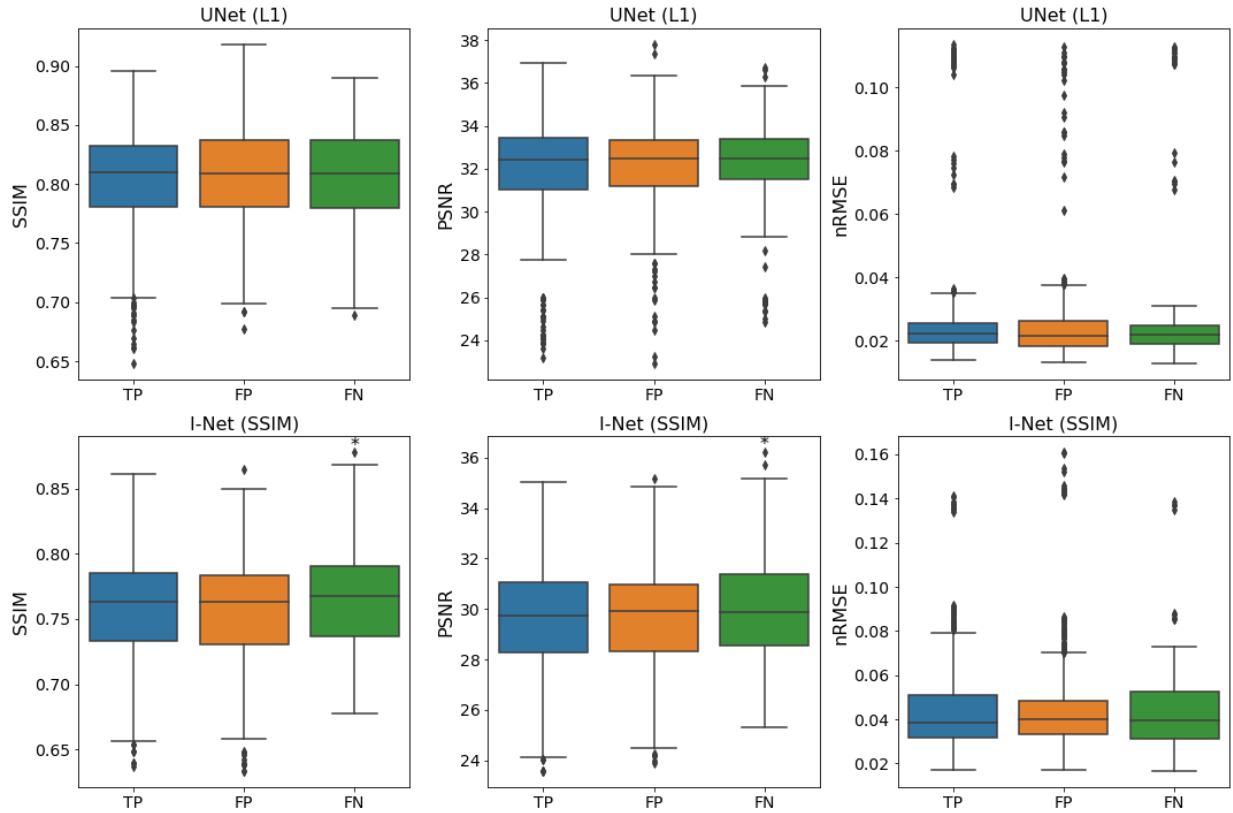

The comparison of mean slice-based SSIM with detection performance. Each slice was classified as having true positive (TP), false positive (FP), or false negative (FN) predictions. A predicted box with a confidence score > 0.70 and IoU > 0.20 was considered a TP. If a slice had multiple types of predictions, its SSIM was included in each of the three groups. One-way ANOVA tests, followed by paired t-tests, were used to determine significant differences in means, marked with asterisks. The plot displays results for the two best-performing reconstruction models based on classic SSIM, with additional plots shown in Figure E2 (supplement). The results revealed no specific pattern indicating that classic SSIM significantly influenced TP, FP, or FN predictions. Among the 14 models analyzed, eight showed significantly higher mean PSNR for the FP group compared to TP and FN, while seven models exhibited lower nRMSE for the FP group.

Figure 6: Relationship Between Changes in Image Quality and Detection Outcomes

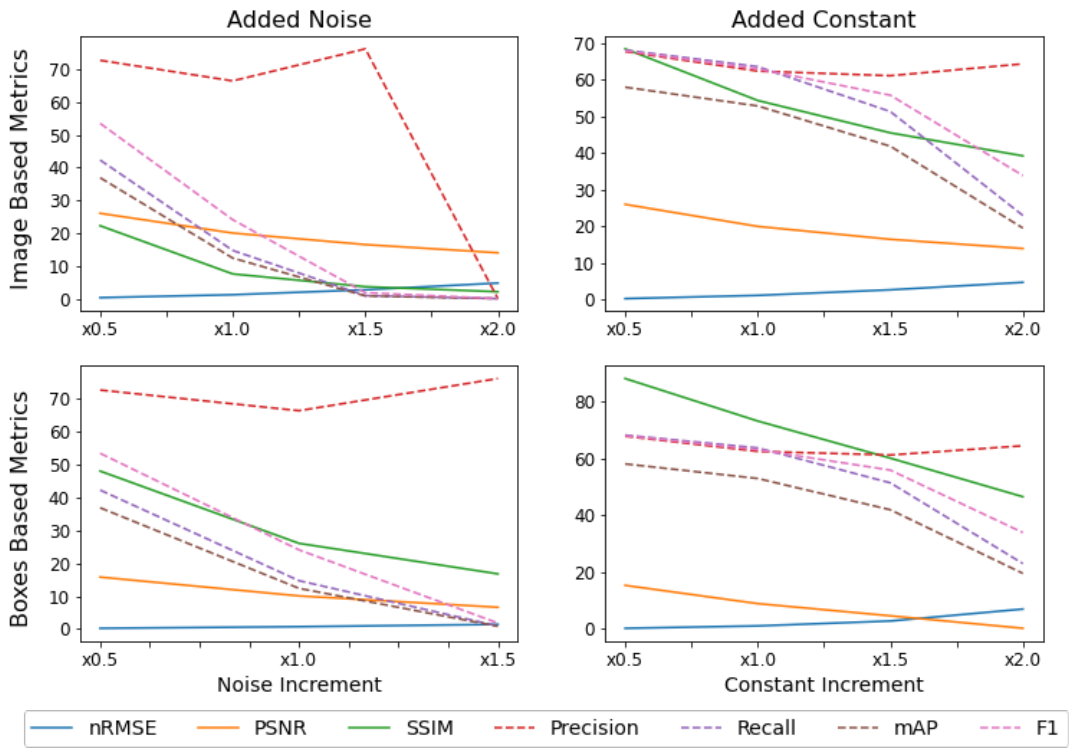

The graphs illustrate the impact of progressively significant artificial negative changes to image quality on detection performance. For each increase in noise, PSNR and SSIM were maintained constant, yet SSIM scores were lower with random noise and significantly higher with the addition of a positive noise (referred to as "constant"). In both scenarios of random noise and constant addition, the decline in detection metrics, aside from precision, strongly correlates with both image-based and box-based SSIM ($r<0.01$, $p<0.05$). Notably, with higher image quality (Added Constant column), detection performance decreases more gradually compared to SSIM, whereas for lower SSIM values (Added Noise column), the reduction in detection metrics is pronounced. This is consistent with other observations that detection performance is less impacted by SSIM variations at higher image qualities, suggesting a threshold beyond which image quality begins to significantly affect the detector's performance.

# Appendix E1

**Anomaly Detection Training Details**

**Normalization.** Normalization was performed in two steps. First, percentile normalization without quality loss was applied to each 3D volume to bring pixel intensity to the standard distribution. A transform called "ScaleIntensityRangePercentiles" from MONAI version 1.0.0 (https://docs.monai.io/) was used with the following setup: lower = 0, upper = 1, clip = False, relative = False. Next, mean and standard deviation were calculated across the whole dataset and supplied as hyperparameters during Faster R-CNN training, with corresponding values in Table E2.

**Data fractionation and upsampling.** Each epoch was trained on 20% of data randomly selected using WeightedRandomSampler from Torch dataloader utils. Our experiments showed a significant increase in training speed with no effect on detection performance results. Additionally, slices with anomalies were upsampled according to weights in Table E1. The sole purpose was to counteract the class imbalance since it was noted during the experimentation phase that least represented anomalies, such as lateral anterior horn lesions, were the most difficult to detect. In fact, a multiclass detection model was trained in one iteration to determine the worst-performing labels, and those labels got the higher weights. Other labels were upsampled with lower coefficients since we still had a much higher number of anomaly-free than labeled slices.

**Data augmentation.** Albumentations version 1.3.0 (https://albumentations.ai/) was utilized to create the augmentation pipeline. Additionally, a custom function was implemented, that randomly shifted a bounding box along the X and Y axes within specified range. This augmentation technique was shown to be the single most important for object detection performance optimization when compared to other standard procedures (21). The algorithm of the augmentation protocol is shown in Table E2.

Table E1: Meniscal Anomaly Labels and Upsampling Weights

| Meniscal compartment | Anomaly boxes count | Upsampling Weight |
|---|---|---|
| Medial meniscal body | 2329 | 2 |
| Medial posterior horn | 8337 | 2 |
| Medial anterior horn | 1084 | 2 |
| Lateral meniscal body | 942 | 3 |
| Lateral posterior horn | 2347 | 3 |
| Lateral anterior horn | 3020 | 3 |
| **Total** | **18059** | - |

Table E2: Object Detection Training Parameters

| Parameters | Specifications |
|---|---|
| Mean/std normalization | Mean, 0.07991; std, 0.07168 |
| Data augmentation | OneOf([<br>   AffineBBoxOnly(translate_px=(-5, 5), scale=(0.8, 1.4), p=0.6),<br>   ShiftScaleRotate(shift_limit=0.0001, scale_limit=0.0001, rotate_limit=15, p=0.6),<br>   RandomCrop(width=400, height=400, p=0.8)<br>], p=0.66),<br>Cutout(num_holes=16, p=0.8) |
| Class ModelParams | num_classes: int=2<br>pretrained_backbone: bool=True<br>min_size: int=800<br>max_size: int=1333<br>image_mean: List[float] = field(default_factory=default_image_mean)*<br>image_std: List[float] = field(default_factory=default_image_std)*<br># RPN parameters<br>rpn_anchor_generator: Optional[AnchorGenerator]=None<br>rpn_head: Optional[RPNHead]=None<br>rpn_pre_nms_top_n_train: int=2000<br>rpn_pre_nms_top_n_test: int=1000<br>rpn_post_nms_top_n_train: int=2000<br>rpn_post_nms_top_n_test: int=1000<br>rpn_nms_thresh: float=0.7<br>rpn_fg_iou_thresh: float=0.7<br>rpn_bg_iou_thresh: float=0.3<br>rpn_batch_size_per_image: int=256<br>rpn_positive_fraction: float=0.5<br>rpn_score_thresh: float=0.0<br># Box parameters<br>box_roi_pool: Optional[MultiScaleRoIAlign]=None<br>box_head: Optional[TwoMLPHead]=None<br>box_predictor: Optional[FastRCNNPredictor]=None<br>box_score_thresh: float=0.05<br>box_nms_thresh: float=0.5<br>box_detections_per_img: int=100 |
| | box_fg_iou_thresh: float=0.5<br>box_bg_iou_thresh: float=0.5<br>box_batch_size_per_image: int=512<br>box_positive_fraction: float=0.25<br>bbox_reg_weights: Optional[List[float]]=None |
| Optimizer and Scheduler | SGD<br>   learning_rate = 0.01<br>   momentum = 0.9<br>   weight decay = 0.0005<br>Scheduler StepLR<br>   step size = 3<br>   gamma = 0.1 |
| Monitor metric | torchmetrics.detection.mean_ap.MeanAveragePrecision() |
| Training parameters | Dataloader<br>   num_workers = 4<br>   epoch_fraction = 0.2<br>Trainer<br>   strategy = 'dp'<br>   accelerator = "gpu"<br>   devices = 2<br>   max_epochs = 30<br>   batch_size = 16 |

Table E3: Relationship Between Image Quality Changes and Detection Performance

| # | Model Name | Recon Image-Based Metrics | | | Detection Metrics | | | |
|---|---|---|---|---|---|---|---|---|
| | | nRMSE | PSNR | SSIM | Precision | Recall | mAP | F1 |
| 1 | Noise x2.0 | 4.74 ± 1.94 | 13.98 ± 0.00 | 2.10 ± 0.71 | 0.00 | 0.00 | 0.00 | 0.00 |
| 2 | Const x2.0 | 4.74 ± 1.94 | 13.98 ± 0.00 | 39.25 ± 4.75 | 64.35 | 22.99 | 19.54 | 33.87 |
| 3 | Noise x1.5 | 2.70 ± 1.11 | 16.48 ± 0.00 | 3.64 ± 1.16 | 76.19 | 0.92 | 0.82 | 1.81 |
| 4 | Const x1.5 | 2.70 ± 1.11 | 16.48 ± 0.00 | 45.53 ± 5.01 | 61.12 | 51.34 | 41.88 | 55.80 |
| 5 | Noise x1.0 | 1.18 ± 0.49 | 20.00 ± 0.00 | 7.53 ± 2.17 | 66.41 | 14.70 | 12.37 | 24.07 |
| 6 | Const x1.0 | 1.18 ± 0.49 | 20.00 ± 0.00 | 54.36 ± 5.18 | 62.38 | 63.59 | 52.87 | 62.98 |
| 7 | Noise x0.5 | 0.30 ± 0.12 | 26.02 ± 0.00 | 22.23 ± 5.06 | 72.69 | 42.26 | 36.88 | 53.45 |
| 8 | Const x0.5 | 0.29 ± 0.12 | 26.02 ± 0.00 | 68.45 ± 4.95 | 67.71 | 68.11 | 57.99 | 67.91 |

Note. nRMSE – normalized root mean squared error. PSNR – peak signal-to-noise ratio. SSIM – structural similarity index, reported in %. All detection metrics are reported in %. mAP – mean average precision. F1 – F1 score. Noise x2.0 – random noise was multiplied by 2.0 and added to the image to generate low SSIM. Const x2.0 – an absolute value of random noise with the same multiplication factor was added to the image to generate high SSIM. For each noise increment, nRMSE and PSNR were held constant. Object detection model failed to predict any anomaly on the lowest quality test set (Noise x2.0).

Figure E1: Boxes-based Metric Calculation Algorithm

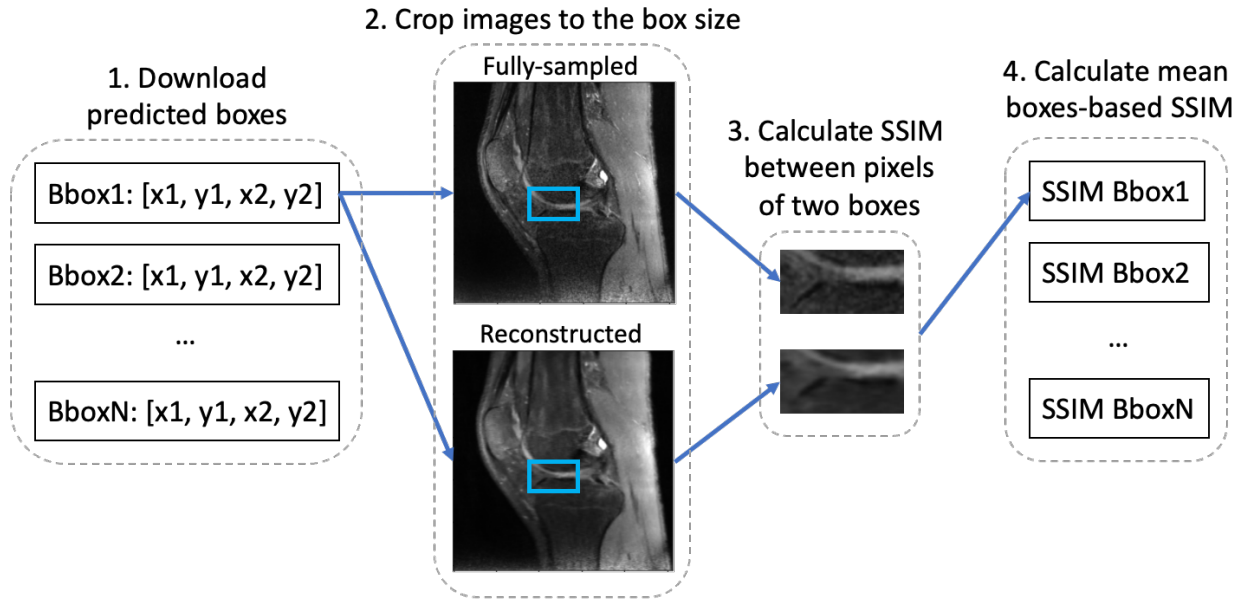

The boxes-based mean SSIM was computed and compared with SSIM obtained using the classic image-based calculation method. For each test set, predictions were acquired and filtered with a score confidence threshold of 0.70. Both fully-sampled and reconstructed images were cropped to match the size of a predicted box. SSIM was then calculated within the cropped regions and averaged across all predicted boxes to derive a comprehensive metric. Box-based PSNR and nRMSE were computed following the same approach. This method aimed to investigate whether box-based metrics exhibit heightened sensitivity to clinically significant image features, such as anomalies and lesions, given the presumed stronger association between predicted boxes and these features.

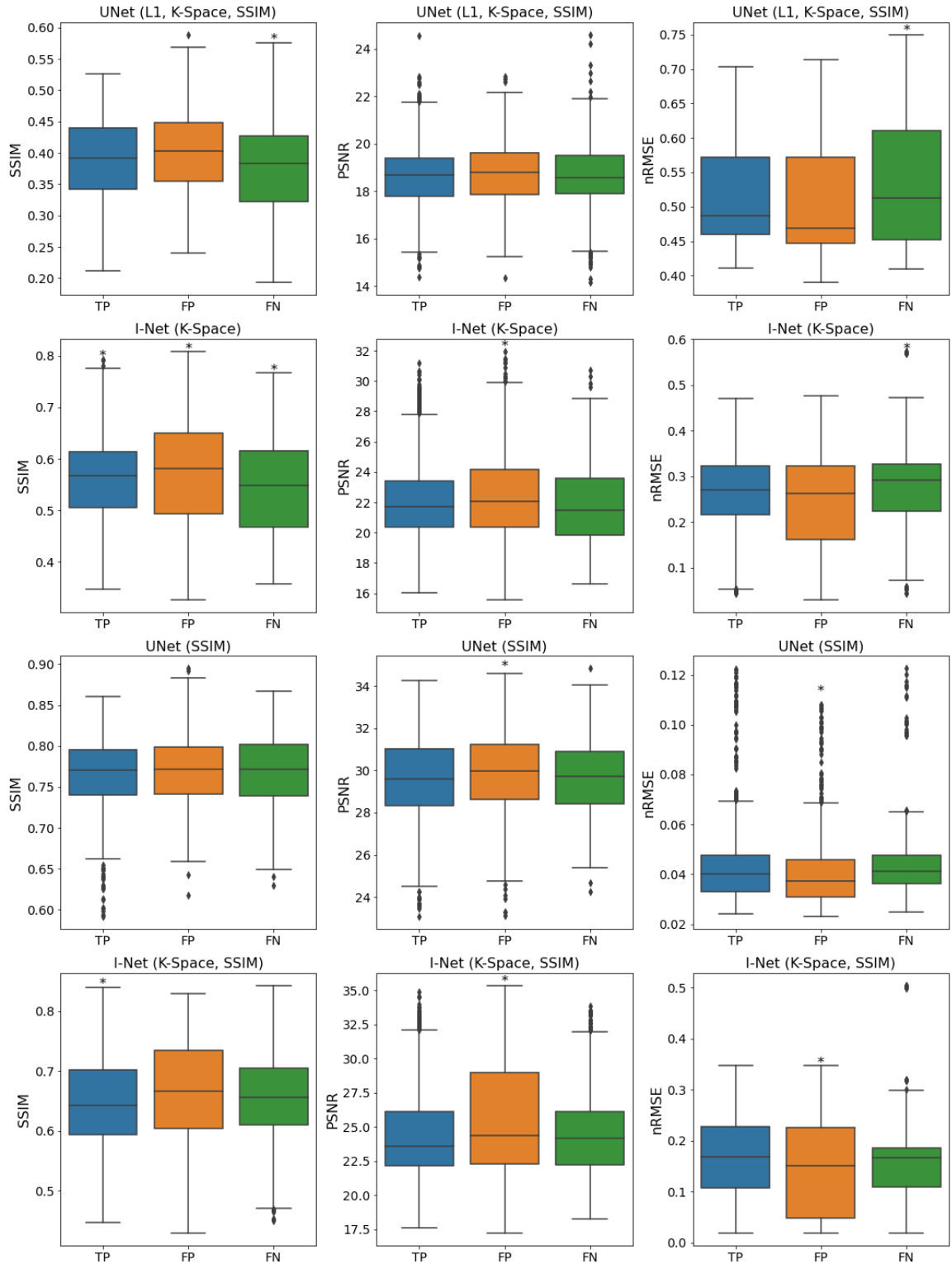

Figure E2: Standard Reconstruction Metrics and Detection Performance

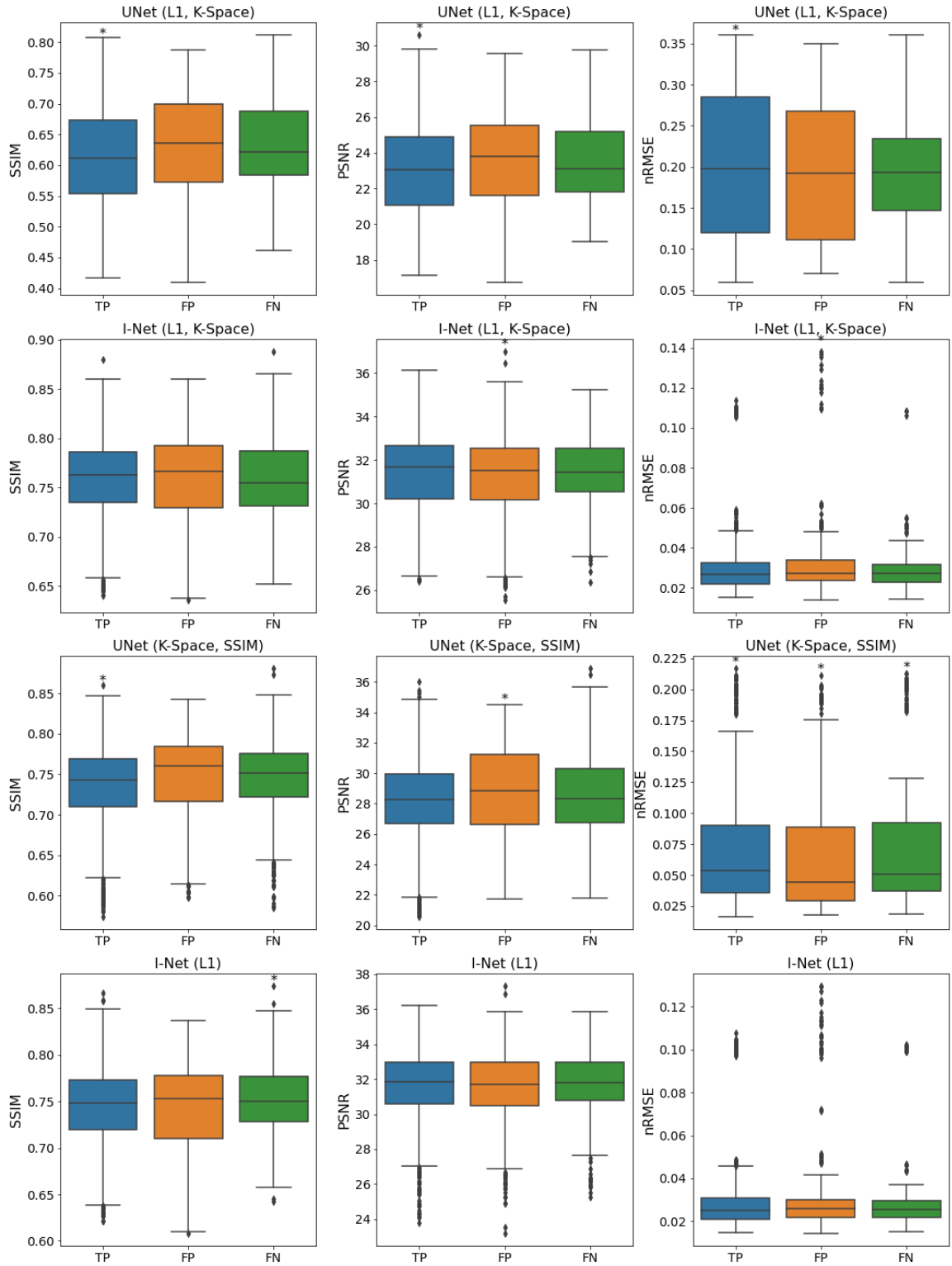

Figure E2 (cont.): Standard Reconstruction Metrics and Detection Performance

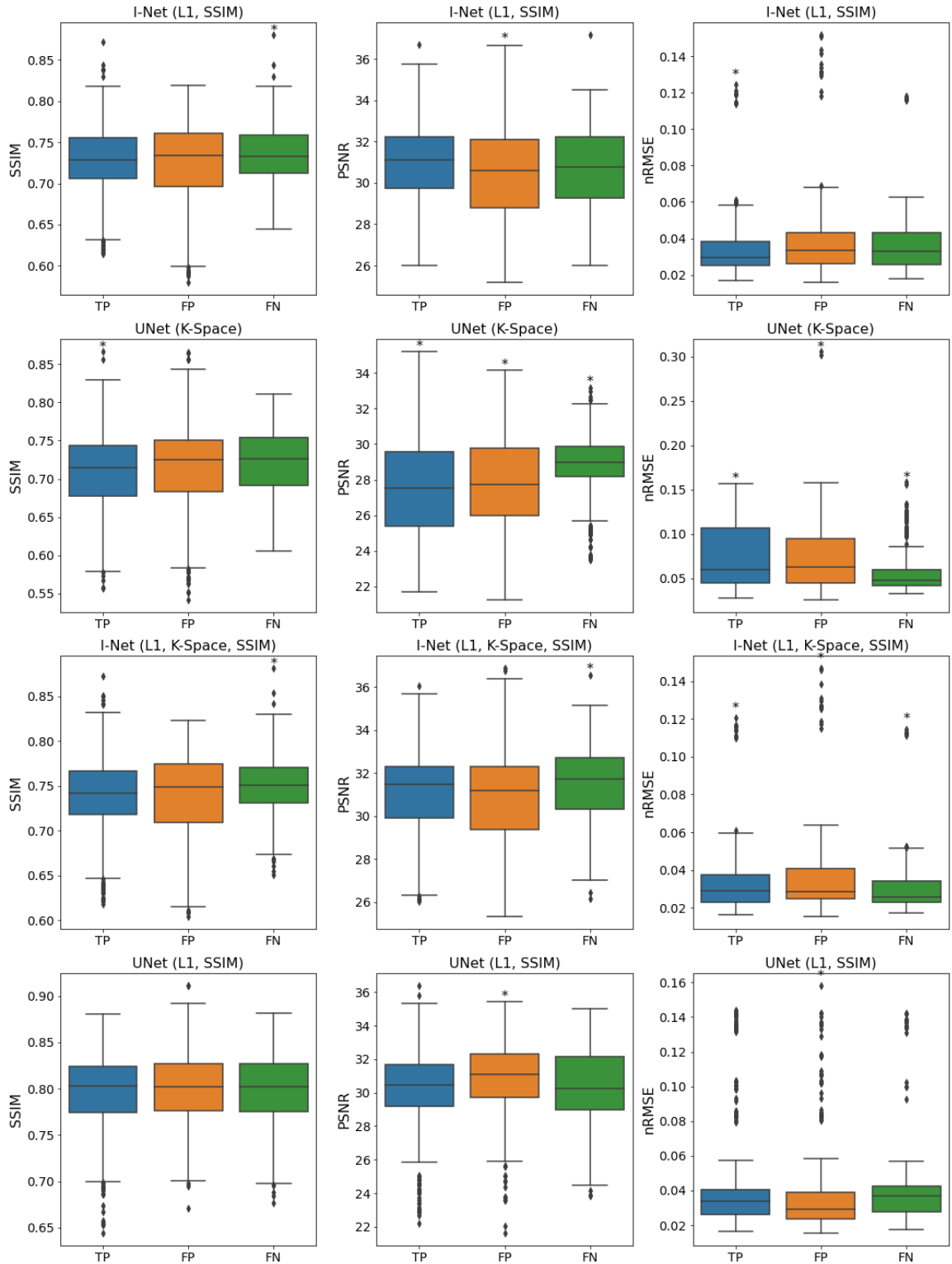

Figure E2 (cont.): Standard Reconstruction Metrics and Detection Performance